\documentclass[%
 reprint,
superscriptaddress,
amsmath,amssymb,
aps,
prl,
]{revtex4-2}

\usepackage{graphicx}% Include figure files
\usepackage{dcolumn}% Align table columns on decimal point
\usepackage{bm}% bold math
\usepackage{afterpage,lipsum}
\usepackage{xcolor}
\begin{document}

\preprint{APS/123-QED}

\title{An Optimal Seed Reveals the Dynamical Route to Spontaneous Splitting of Turbulent Puffs}

% === 严格保持作者顺序 ===
\author{Chihang Liu}
\email{These authors contributed equally to this work.}
\affiliation{
  Department of Mechanics, 
  School of Mechanics and Engineering Science, 
  Peking University, 
  Beijing, 100871, China
}

\author{Hao Ni}
\email{These authors contributed equally to this work.}
\affiliation{ 
  College of Engineering, 
  Peking University, 
  Beijing, 100871, China
}

\author{Jiashun Guan}
\email{jiashun.guan@upm.es}
\affiliation{
  Department of Mechanics, 
  School of Mechanics and Engineering Science, 
  Peking University, 
  Beijing, 100871, China
}
\affiliation{
  School of Aeronautics, 
  Universidad Politécnica de Madrid, 
  Madrid, 28040, Spain.
}

\date{\today}% It is always \today, today,

\begin{abstract}

The localized turbulence in pipe flow, known as puffs, is sustained by their proliferation. This process is crucial to the subcritical laminar-turbulent transition, a century-old fundamental problem related to a wide range of natural and engineered systems. Here, we leverage the nonlinear nonmodal analysis of a reduced-order model to capture the full Navier–Stokes dynamics of the splitting process: from the splitting seed through the \textit{edge} state to successful splitting, even when the pre-\textit{edge} evolution is obscured by the puff's intrinsic chaotic fluctuations. Moreover, the optimized evolution provides a splitting-progress coordinate to frame a phase-space representation of the proliferation, resolving the temporal desynchronization among chaotic realizations and thereby enabling dynamically aligned ensemble averaging to uncover the underlying splitting dynamics. The splitting seed is characterized by a strong velocity inflection, and its energy threshold exhibits a power-law scaling with the Reynolds number, giving the optimal solution direct physical consistency in spontaneous puff splitting.

\end{abstract}

\maketitle

%\section{Introduction}
As one of the best-known canonical flows in natural and engineered systems, pipe flow was pioneered by Hagen and Poiseuille in the first half of the 19th century \cite{Hagen1839, Poiseuille1840}. It later served as a prototype in Reynolds' pipe flow experiments \cite{Reynolds1883_PT} and paved the way for our understanding of laminar-turbulent transition. Over nearly $190$ years, experimental and numerical observations \cite{Hof2006, WillisKerswell2007, PeixinhoMullin2007, AvilaWillisHof2010, avila2011, NemotoAlexakis2021,Mullin2011} and theoretical interpretations \cite{barkley2015rise, ShihHsiehGoldenfeld2016, Lemoult2024, guan2025,guan2025a} have revealed that the subcritical transition in linearly stable pipe flow is governed by transient turbulent puffs at moderate Reynolds numbers ($Re$), and that the turbulent state is sustained when the proliferation of puffs outweighs their decay \cite{avila2011}. These two processes are both statistically memoryless, and their lifetimes, probability distributions, and scalings with $Re$ have been investigated systematically \cite{AvilaBarkleyHof2023,Manneville2016}.  

Zooming in from ensemble statistics to individual events, the dynamics of puff proliferation, i.e., how a single puff splits into two, are crucial for understanding the onset of sustained turbulence. The critical state along this transition has been captured by a bisection method and identified as an \textit{edge}, and the process has been phenomenologically described as the slug-gap-split mechanism \cite{frishman2022b, Svirsky2026}. 

While the \textit{edge} organizes the observed puff-splitting transition path, how such a trajectory emerges from the intrinsic chaotic fluctuations and subsequently develops toward proliferation through nonlinear dynamics remains unresolved. Conceptually, the onset of this splitting pathway can itself be viewed as an earlier \textit{edge}-like critical point, separating intrinsic fluctuations that return to the one-puff dynamics from those that develop toward \textit{edge} and successful splitting. Unlike the identified \textit{edge} \cite{Svirsky2026}, however, its earlier trajectories are embedded within the chaotic fluctuations and cannot be directly identified by bisection.   The intrinsic chaos further leads to nonmonotonic progression and temporal desynchronization among splitting realizations, covering their shared dynamical evolution and defying a consistent ensemble description.

Here, we identify the splitting seed of puff within the Barkley model \cite{barkley2015rise,barkley2016theoretical} by nonlinear nonmodal analysis and construct an indicator factor based on inner products of flow profiles to track the puff-splitting trajectory in phase space. Combining deterministic optimized states and direct numerical simulation results, we further reveal the optimal splitting path, characteristic flow structures, and energy threshold governing the puff splitting process.

%\section{Methods}
%NS方程描述湍塞，进而引出SI验证部分的寿命
The incompressible Navier-Stokes equations are numerically solved by OpenPipeflow \cite{willis2017} to simulate the puff dynamics in pipe flow with length $32\pi D\approx100D$ at $Re=2230$. The velocity $(u_x,u_\rho,u_\theta)$ and pressure fields were spatially discretized in cylindrical coordinates $(x,\rho,\theta)$. The Reynolds number $Re$ is defined as $Re= U_b D/\nu$, where $\nu$ is the kinematic viscosity, $D$ is the pipe diameter, and $U_b$ is the bulk velocity in the streamwise direction. No-slip conditions are applied at the side walls, and periodicity is imposed in the streamwise direction. Setups of direct numerical simulations (DNS) are detailed in the SI.
%引入Barkley模型，刻画puff
The Barkley model \citep{barkley2015rise,barkley2016theoretical} captures the key characteristics of puff decay and splitting and has been widely investigated for puff dynamics \citep{Frishman2022,frishman2022b}. Here, the model is simulated with a spectral method whose details are given in the SI.
% 雷诺数转换
To establish a correspondence between the Navier–Stokes dynamics and the Barkley model, the Reynolds number ($Re$) and the Reynolds-like parameter ($r$), which will also be denoted as the Reynolds number, are converted by the linear mapping $r=\frac{r_s-r_c}{Re_s-Re_c}(Re-Re_c)+r_c$, where the subscripts (s) and (c) represent, respectively, the puff–slug boundary ($Re_s=2250$ and $r_s=0.72$), and the onset of the saddle-node bifurcation in the puff dynamics ($Re_c=1790$ and $r_c=0.425$) \cite{barkley2015rise,barkley2016theoretical,Frishman2022,guan2025}. Here, we consider $Re=2230$, corresponding to $r\approx0.7$.
%引出NNO方法，寻找split seed
Nonlinear nonmodal optimization (NNO) has been established as a tool for seeking the minimal seed of transition to turbulence starting from a laminar state \citep{kerswell2014optimization,kerswell2018nonlinear} and extreme events in turbulence \citep{farazmand2017,Ciola2023}. Here, it is used in the Barkley model to identify the optimal state that could trigger puff splitting. However, the puff could remain turbulent for a long time, and the basic state of optimization is set as a stable single-puff stage in the noise-free Barkley model. The maximal disturbance energy at target time $T$ is set as the objective function. The optimization seeks the minimum-energy initial perturbation that successfully triggers puff splitting from the basic state. This perturbation is termed the splitting seed. More settings about nonlinear nonmodal analysis are shown in the SI.

%\section{proper noun}
%\textbf{minimal splitting seeds}: 
%The initial fields corresponding to the minimum perturbation energy capable of triggering puff splitting at a fixed Reynolds number $r$

%\textbf{indicator factor $I$}: 
%inner product calculations between the optimized $q$ and $u$ profiles at different time instants and the corresponding DNS profiles at each moment

%\section{Results}
%NNO结果
\begin{figure} 
    \centering
    \includegraphics[width=0.95\linewidth]{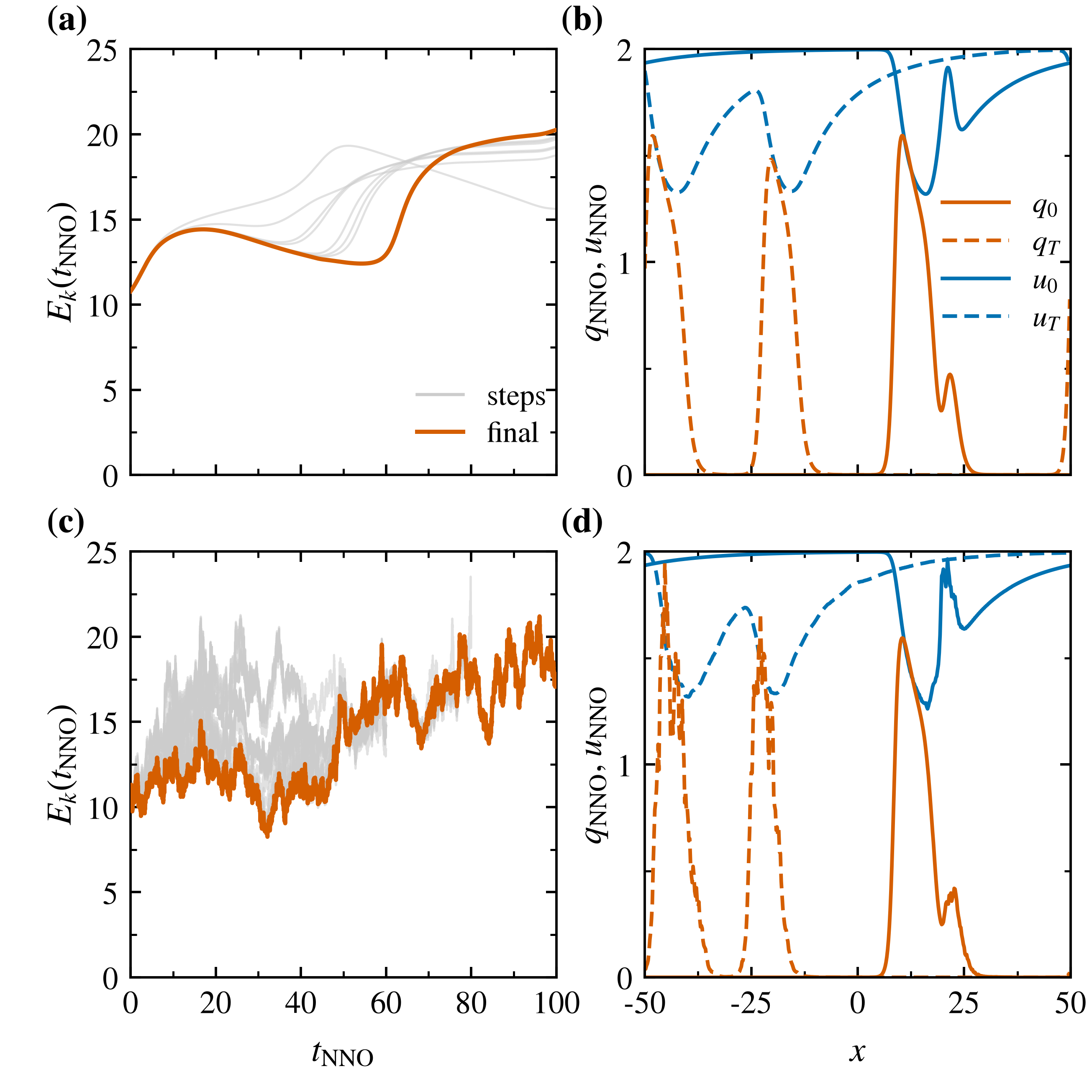}
    \caption{NNO results of the Barkley model without and with noise. (a) and (c) are the temporal evolution of disturbance kinetic energy $E_k$ from the optimization of the Barkley model without and with a noise term, respectively. The orange lines denote results from the final optimization step, while the gray lines correspond to intermediate optimization steps. (b) and (d) are profiles of the initial field ($t_\mathrm{NNO}=0$, solid lines) and the target‑time field ($t_\mathrm{NNO}=T$, dashed lines) obtained from the optimization of the Barkley model without and with a noise term, respectively. The blue lines represent the centerline velocity $u$, and the orange lines stand for the local turbulence intensity $q$, which physically represents an integration of the turbulent fluctuations in cross-section in the Barkley model.}
    \label{fig:NNO}
\end{figure}

\begin{figure*}
    \centering
    \includegraphics[width=\linewidth]{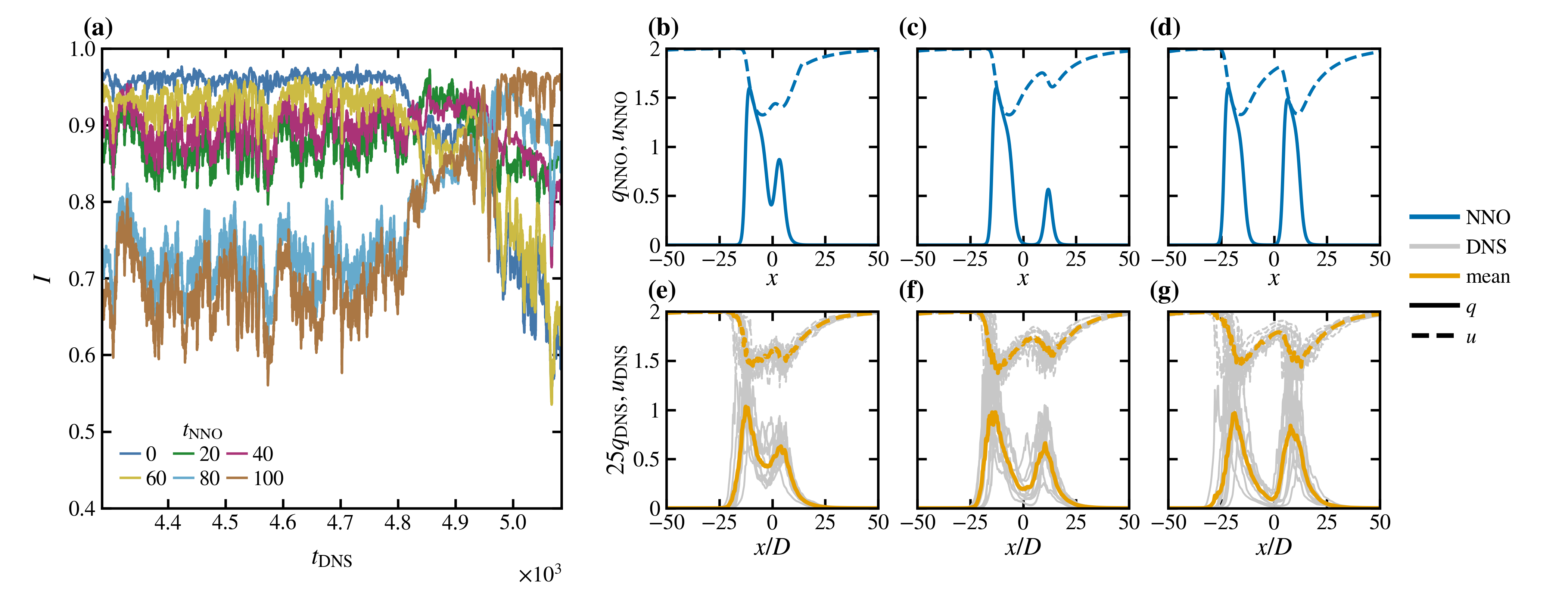}
    \caption{Comparison of NNO profiles and their corresponding DNS evolution. (a) The indicator factor $I$ varies with DNS time obtained from optimization in the deterministic Barkley model. Curves of different colors correspond to profiles from different optimization instants. The horizontal axis denotes the DNS evolution time, and the vertical axis represents the magnitude of the indicator factor $I$. (b), (c), and (d) show optimized profiles at $t_\mathrm{NNO}=10, 60, 100$, while panels (e), (f), and (g) present the corresponding DNS profiles.}
    \label{fig:DNS}
\end{figure*}

Different from full Navier-Stokes dynamics, where intrinsic turbulent fluctuations cannot be selectively removed, the stochastic term in the Barkley model can be turned off, providing a controlled setting to examine whether the optimal splitting pathway is rooted in the deterministic dynamics or qualitatively governed by the chaotic fluctuations. As illustrated by the red curves (final optimized states) in Fig.~\ref{fig:NNO}(a) and (c), the temporal evolution of disturbance kinetic energy $E_k=\frac{1}{2}[(q^2+(U_0-u)^2]$ is nearly identical for both cases, where $q$, $u$ and $U_0$ are the local turbulence intensity, centerline streamwise velocity and laminar centerline streamwise velocity in Barkley model, respectively. Starting from a low energy level around $10$, the kinetic energy rises rapidly and then saturates at a plateau, referred to as the edge stage \citep{Chantry2014,frishman2022b,guan2025,Svirsky2026}. After this ``bottleneck'' plateau, the kinetic energy reaches its maximum at the target optimization time $T$, and the flow field exhibits signatures of two individual puffs, indicating puff splitting, as shown in Fig.~\ref{fig:NNO}(b) and (d).
In the optimized initial conditions ($t_\mathrm{NNO}=0$), the distribution of $q$ develops a small downstream peak, while the corresponding downstream position in the $u$‑profile features velocity inflection points close to the laminar solution. This observation bears resemblance to the inflection points detected at the turbulent‑laminar interfaces in real pipe‑flow experiments \citep{hof2010}. Such velocity inflection points provide an energy source for the puffs and facilitate their splitting. 
These findings suggest that, within the Barkley model, the dynamical characteristics of the optimized final states remain essentially consistent regardless of whether noise is incorporated. Consequently, the presence or absence of the noise term does not alter the underlying dynamics of the puff‑splitting.
%选择非噪声B-Model优化的原因
The following analysis therefore focuses on optimization of the noise-free Barkley model, which retains the deterministic dynamics underlying chaotic puff splitting while providing an efficient system for NNO that would be prohibitively expensive in the full Navier–Stokes system.

%NNO和DNS对比

The optimization results without the noise term are adopted for comparison with the puff splitting process in real pipe flow. An indicator ($I$) is defined by projecting the instantaneous $q_\mathrm{DNS}$ and $u_\mathrm{DNS}$ profiles onto the optimized evolution, using the streamwise-rescaled optimized profiles at different stages of the splitting process as reference states. Here, $q_\mathrm{DNS}=\frac{1}{2}\int_S(u_x^2+u_{\rho} ^2+u_\theta^2)dS$, where $S$ denotes the cross-section of the pipe, and $(u_x,u_\rho,u_\theta)$ is the three components of disturbance velocity at centerline. And $u_\mathrm{DNS}$ is the streamwise velocity at the centerline. Details of the definition of $I$ are shown in the SI.
These two quantities also have corresponding state variables in the Barkley model, denoted by $q_\mathrm{NNO}$ and $ u_\mathrm{NNO}$. The indicator factor $I$ is a function of the DNS evolution time $t_\mathrm{DNS}$ and the optimization time $t_\mathrm{NNO}$. For a given instant in the optimization procedure, the optimized profile $(q_\mathrm{NNO}, u_\mathrm{NNO})$ at that instant can be obtained. Substituting this optimized profile and the DNS‑evolved profile $(q_\mathrm{DNS}, u_\mathrm{DNS})$ into the aforementioned definition of $I$ yields the temporal evolution sequence of $I$.
Figure~\ref{fig:DNS}(a) illustrates the temporal evolution of the indicator factor $I$ for a typical splitting process evaluated from optimized profiles at different temporal instants. All curves exhibit steady fluctuations in the early stage, followed by abrupt variations that correspond to the pre-\textit{edge}, \textit{edge}, and successful splitting. Larger values of the inner-product-based $I$ indicate higher consistency between flow profiles, suggesting that the optimized evolution captures the flow structure progression observed during DNS-resolved puff splitting. The temporal order of these peaks characterizes the sequential emergence of specific flow structures throughout the splitting process.

Long before the successful splitting, the DNS puff remains predominantly aligned with the early stages of the optimized trajectory, while showing little similarity to its later stages. This pattern undergoes a reorganization as splitting develops: the projection onto early optimal states weakens (Fig.~\ref{fig:DNS}(a) for $t_\mathrm{NNO}=0,20$), whereas that onto progressively later states rises and peaks in temporal order (Fig.~\ref{fig:DNS}(a) for $t_\mathrm{NNO}=40,60,80,100$). The resulting transfer of similarity along the optimized trajectory indicates that the DNS trajectory evolves through the same dynamical sequence, despite the underlying chaotic fluctuations. The redistribution of projection among successive optimal states therefore provides a dynamical signature of the DNS trajectory progressing along the optimized splitting route. It should be noted that, although the optimization identifies the minimum-energy route to splitting, a spontaneous chaotic fluctuation need not pass through the optimal initial state itself. Rather, natural trajectories may enter the vicinity of the optimized route from different states and subsequently be captured by the same dynamical progression toward splitting.

The profiles between the DNS and optimal splitting path show a strong consistency. Figure~\ref{fig:DNS}(b)‑(d) and (e)‑(g) show profile comparisons at the peak-$I$ time instants from optimization and DNS, respectively. Detailed calculations are provided in the SI. The three characteristic instants correspond to the stages before, during, and after the \textit{edge} state, capturing the evolutionary pathway of the puff from the vicinity of splitting seed to successful splitting. At $t_\mathrm{NNO}=10$, as shown in Fig.~\ref{fig:DNS}(b) and (e), the small downstream peak in the $q$ profile begins to grow while remaining largely consistent with that of the splitting seed in Fig.~\ref{fig:NNO}. Meanwhile, the upstream portion of the $u$ profile undergoes substantial modification: the velocity inflection point of the splitting seed transforms into a profile resembling a turbulent centerline profile, and the perturbation induced by this velocity-profile transition propagates downstream. This accounts for the initial energy growth of the puff after energy extraction from the velocity inflection point within the splitting seed. This mechanism shares similarities with the slug-gap-split mechanism adopted in prior work to interpret puff splitting via flow fluctuations \citep{frishman2022b}. The present analysis focuses on the early stage. It further clarifies that the emergence of the gap originates from the strong velocity inflection point embedded in the splitting seed, which drives the formation of the turbulent region downstream of the puff. The puff subsequently evolves into the edge state \cite{frishman2022b,Svirsky2026}, as illustrated in Fig.~\ref{fig:DNS}(c) and (f), before finally reaching the fully split state with two coexisting puffs, shown in Fig.~\ref{fig:DNS}(d) and (g).

%DNS时的相空间演化
\begin{figure*}
    \centering
    \includegraphics[width=\linewidth]{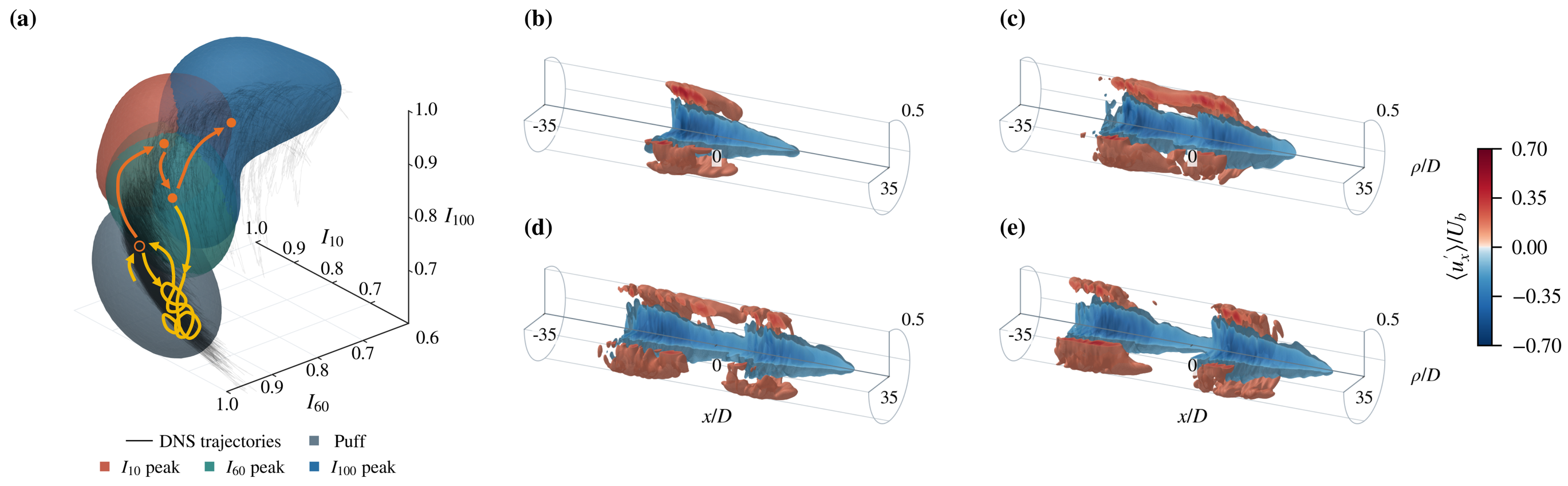}
    \caption{Phase-space evolution and characteristic flow structures along the puff-splitting path. (a) DNS trajectories and probability regions in the three-dimensional phase space $(I_{10},I_{60},I_{100})$, where the subscript denotes $t_{\rm NNO}$. Gray curves show DNS trajectories. Gray, red, green, and blue surfaces represent the equilibrium puff state and states near the respective $I_{10}, I_{60}$, and $I_{100}$ peaks. Probability regions are estimated using Gaussian kernel density estimation (KDE) with bandwidth $\sigma_{\mathrm{kernel}}=0.05$ and enclose $68\%$ of the probability mass, with equal weights assigned to the $11$ cases. The equilibrium puff state (the gray surface) comprises $100$ DNS time units ($D/U_b$) randomly selected from quasi-equilibrium puff time-series per case, while peak-associated states (the color surfaces) are taken within $\pm1.25$ DNS time units ($D/U_b$) of each peak. Filled orange circles mark the expected coordinates under the probability density functions (pdf) in KDE represented by the red, green, and blue surfaces. The open orange circle marks the phase-space locations of the $t_{\mathrm{NNO}}=0$ peaks for an individual splitting event (DNS case8 listed in Tab.~\ref{tab} in the SI). The yellow arrows diagram the chaotic evolution within the equilibrium puff state, while the orange arrows diagram an optimal splitting path. Panels (b)-(e) show the ensemble-averaged streamwise disturbance velocity iso-surfaces, conditioned on the peaks of the indicator \(I\) associated with the puff state and the optimal splitting path at $t_{\mathrm{NNO}}=10, 60, 100$, respectively. The ensemble averages are computed using 4 of the 11 DNS cases; see Tab.~\ref{tab} in the SI.}
    \label{fig:phase_space}
\end{figure*}

Beyond the consistency in profile evolution, the optimal splitting path provides an intrinsic progress coordinate with which the phase-space evolution of puff splitting can be reconstructed, allowing chaotic realizations to be dynamically synchronized along a common transition route. The indicator factor $I$ evaluated from optimized profiles at these three instants serves as the three coordinates in Fig.~\ref{fig:phase_space}(a), yielding the phase-space trajectory of the puff-splitting process. Regions with different colors mark the locations corresponding to distinct puff states within the phase space.
When the puff remains as a single entity, the flow state stays predominantly within the gray region, fluctuating inside this basin without escape, diagrammed by yellow trajectories in Fig.\ref{fig:phase_space}a. As splitting commences from an \textit{edge}-like critical point (exemplified by the open orange circle in Fig.\ref{fig:phase_space}a), the trajectory departs from the steady puff region and proceeds toward the red region. At this location, the $(q_\mathrm{DNS},u_\mathrm{DNS})$ profiles bear resemblance to the profiles from the optimal splitting path at $t_\mathrm{NNO}=10$. As illustrated in Fig.~\ref{fig:phase_space}(c), pronounced local relaminarization occurs downstream of the parent puff, which is consistent with the characteristics of the $u$ profiles in Fig.~\ref{fig:DNS}(b)(e). As splitting evolves, the flow reaches the edge state, corresponding to the green region in Fig.~\ref{fig:phase_space}(a) and the dynamically aligned ensemble flow structures shown in Fig.~\ref{fig:phase_space}(d). Eventually, the flow arrives at the state with two separated puffs, marking the successful splitting, as demonstrated by the blue region in Fig.~\ref{fig:phase_space}(a) and the ensemble flow structures in Fig.~\ref{fig:phase_space}(e). The dynamically aligned ensembles, Fig.~\ref{fig:DNS}(c-e), resolve a reproducible three-dimensional structural evolution through elongation, gap formation, and successful splitting, elevating the previously phenomenological ``slug-gap-split'' mechanism \cite{frishman2022b} to an ensemble-resolved dynamical pathway in the full Navier--Stokes system.

%变化Re时的NNO特征
\begin{figure*}
    \centering
    \includegraphics[width=\linewidth]{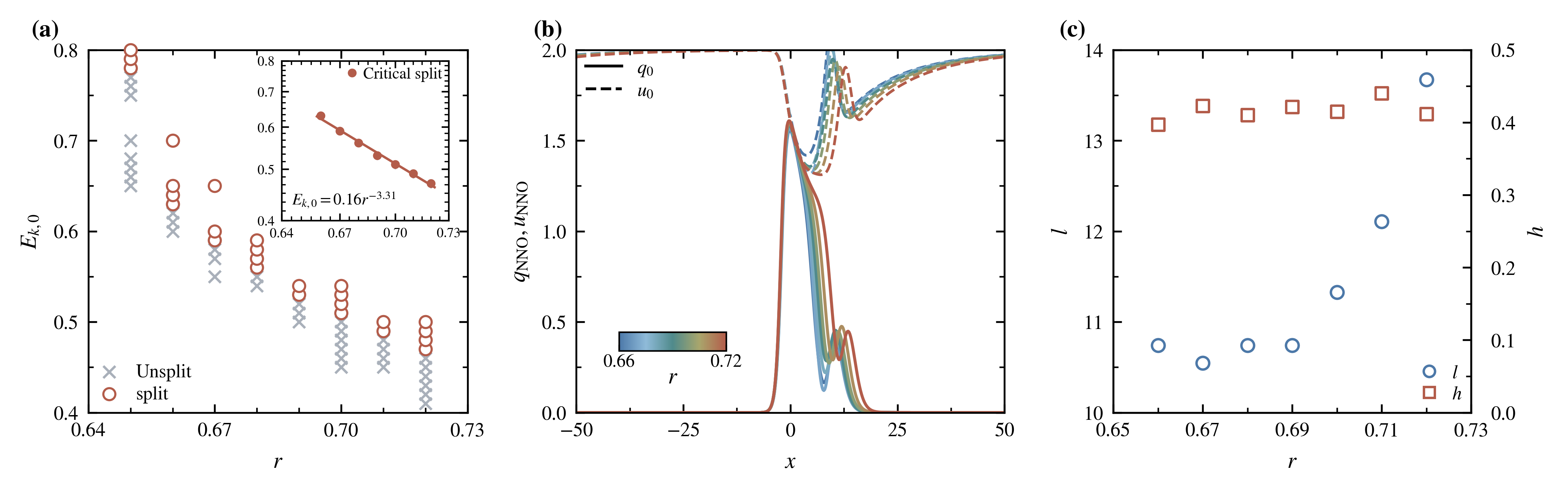}
    \caption{Characteristics of optimal seeds identified by NNO at different Reynolds numbers $r$. (a) The optimization cases without a noise term in the phase space spanned by $r$ and initial perturbation energy $E_{k,0}$. Red circles denote splitting cases, and gray crosses represent non‑splitting cases. The inset shows results plotted on a log‑log scale, where the solid line indicates the fitted power‑law relation \(E_{k,0} \sim r^{-\gamma}\), yielding \(\gamma=3.31\), with a log-space goodness of fit $R^2$ of 0.993. (b) The initial fields corresponding to the minimum perturbation energy capable of triggering puff splitting at a fixed $r$, i.e., the splitting seeds. Solid lines stand for $q$ and dashed lines for $u$. Different colors correspond to splitting seeds under different Reynolds numbers. (c) The variations of perturbation peak height $h$ and peak distance $l$ of the splitting seeds with Reynolds number $r$. The perturbation peak height $h$ is defined as the height of the downstream peak in the $q$ profile after subtracting the single‑puff profile. The peak distance $l$ refers to the distance between the two peaks in the $q$ profile.}
    \label{fig:DiffRe}
\end{figure*}

Having established the correspondence with DNS, we now leverage the noise-free Barkley model to explore how the optimal splitting dynamics vary with the Reynolds number $r$.
Figure~\ref{fig:DiffRe}(a) shows the noise-free optimization results in the parameter plane spanned by the initial perturbation energy and the Reynolds number $r$. A higher Reynolds number $r$ corresponds to lower perturbation energy required to trigger puff splitting, and this dependence approximately follows a power‑law scaling $r^{-3.31}$. Only splitting seeds at $r>0.65$, where the stable turbulent state shows up \citep{Frishman2022}, are included in the scaling fitting. This behavior is reminiscent of the power‑law scaling $E \sim Re^{-\gamma}$ reported for transition energy threshold estimates \citep{Trefethen1993}. In physical pipe flow experiments \cite{avila2011}, puffs undergo spontaneous splitting more rapidly at elevated Reynolds numbers, meaning that splitting occurs more readily. This trend is consistent with the decreasing susceptibility threshold for triggering a splitting event. Note that the perturbations required for splitting originate from the puff's intrinsic chaotic dynamics, whose fluctuation strength remains nearly constant across the puff's Reynolds number regime \cite{guan2025}. As in minimal-seed problems for laminar–turbulent transition \cite{kerswell2018nonlinear}, however, exceeding this energy scale alone is not sufficient: the fluctuation must also acquire an appropriate flow structure configuration, leaving the spontaneous initiation of splitting intrinsically stochastic \cite{jimenez2024}. The differently‑colored curves in Fig.~\ref{fig:DiffRe}(b) illustrate the streamwise profiles of $q_{\mathrm{NNO}}$ and $u_{\mathrm{NNO}}$ for varying $r$. As $r$ increases, the downstream peak of $q_{\mathrm{NNO}}$ shifts progressively downstream, while its amplitude $h$ changes only weakly, as quantified in Fig.~\ref{fig:DiffRe}(c).  Meanwhile, the near‑laminar region ($u_{\mathrm{NNO}}$ near to $2$) in the $u$‑profile shrinks and also moves downstream (Fig.~\ref{fig:DiffRe}(b)). Thus, the decrease in the susceptibility threshold with increasing $r$ is accompanied not simply by a weaker perturbation, but by an increasing localization of the optimal disturbance in the downstream region of the puff. Since the intrinsic fluctuation strength of the puff remains nearly constant over puff's Reynolds-number range \cite{guan2025}, the reduced splitting threshold cannot be simply attributed to stronger background fluctuations, but instead reflects an increased susceptibility to localized downstream fluctuations.

%\section{Conclusion}
In this Letter, nonlinear optimization in the Barkley model identifies an optimal route for puff self-replication, serving as a reduced-order probe to resolve transition dynamics that are prohibitively difficult \cite{Ciola2023} to directly implement in the chaotic Navier–Stokes system. The optimized evolution provides a dynamical coordinate that organizes spontaneous DNS trajectories along the same route, resolves their temporal desynchronization \cite{gome2022}, and enables an ensemble-resolved description of the ``slug-gap-split'' splitting process \cite{frishman2022b,Svirsky2026}. The resulting splitting seed exhibits a velocity inflection \cite{hof2010,song2017} and a Reynolds-number-dependent energy threshold \cite{Trefethen1993}, connecting the optimized route with the physical dynamics of spontaneous puff splitting.

More broadly, our results suggest that reduced-order dynamics \cite{barkley2015rise,barkley2016theoretical,guan2025,Benavides2025,Rowley2017} can provide a tractable window into the underlying dynamics of high-dimensional chaotic systems, extending nonlinear optimization from the search for optimal perturbations to the identification of transition coordinates \cite{kerswell2018nonlinear, Huhn2020, farazmand2017}. At a more fundamental level, this view is also reminiscent of the instanton description of rare events \cite{Gurarie1996,Bouchet2019}, but suggests that preferred transition pathways may arise from the deterministic dynamics of chaos itself.  Returning to the century-old transition problem, the splitting seed and its Reynolds number dependence identified here provide a potential basis for theoretically understanding the super-exponential scaling of puff-splitting lifetimes \cite{avila2011,AvilaBarkleyHof2023}, while the same framework may point toward a more universal description of transition across subcritical shear flows \cite{Tuckerman2020}.

%\tableofcontents

% % for the sample file to show the different styles of references, but authors
% % most likely will not want to use it.
% \nocite{*}

\bibliography{apssamp}% Produces the bibliography via BibTeX.

\clearpage
\onecolumngrid

\appendix

\setcounter{figure}{0}  % 重置图计数器
\renewcommand{\thefigure}{S\arabic{figure}}  % 改为 S1, S2, 
\section{\large Supplementary Information}

\subsection{Perturbation equations of the puff}

We consider the Barkley model for pipe flow~\cite{barkley2015rise,barkley2016theoretical}:
\begin{equation}
\begin{cases}
\dfrac{\partial q_{\rm tot}}{\partial t}
+(u_{\rm tot}-\zeta)\dfrac{\partial q_{\rm tot}}{\partial x}
=f(q_{\rm tot},u_{\rm tot})
+D_q\dfrac{\partial^2 q_{\rm tot}}{\partial x^2}
+\sigma q_{\rm tot}\eta,
\\[10pt]
\dfrac{\partial u_{\rm tot}}{\partial t}
+u_{\rm tot}\dfrac{\partial u_{\rm tot}}{\partial x}
=g(q_{\rm tot},u_{\rm tot}),
\end{cases}
\label{eq:main_system}
\end{equation}
where
\begin{align}
f(q_{\rm tot},u_{\rm tot}) &= q_{\rm tot} \left[ r + u_{\rm tot} - U_0 - (r + \delta)(q_{\rm tot} - 1)^2 \right], \\
g(q_{\rm tot},u_{\rm tot}) &= \varepsilon_1 (U_0 - u_{\rm tot}) + \varepsilon_2 (\overline{U} - u_{\rm tot})\, q_{\rm tot}.
\end{align}
Here, $t$ denotes the evolution time of the Barkley model, which is equivalent to $t_\mathrm{NNO}$ in the main text. $q_{\rm tot}(x,t)$ denotes the local turbulence intensity,
$u_{\rm tot}(x,t)$ denotes the centreline velocity, and $r$ is the
Reynolds-number-like control parameter. $U_0$ is the laminar centreline
velocity and $\overline{U}$ is the bulk velocity.
$\zeta$ sets the
relative advection speed of the turbulent field.
$D_q$ is the effective
streamwise diffusion coefficient.
$\delta$ controls the cubic reaction
term.
$\varepsilon_1,\varepsilon_2$ set the relaxation rates of
$u_{\rm tot}$.
% The multiplicative noise amplitude is denoted by $\sigma$, and
% $\eta(x,t)$ is a zero-mean Gaussian white noise with
% $\langle \eta(x,t)\eta(x',t')\rangle=\delta(x-x')\delta(t-t')$.\nh{Seem inconsistent with the numerical method below. We should change the expression here because code is written in the form below.}
The multiplicative noise amplitude is denoted by $\sigma$, and
$\eta(x,t)$ is a zero-mean Gaussian white noise.
The system is posed on a periodic domain in $x$, and all spatial
integrals below are taken over this domain.
% Unless otherwise stated, we use the Barkley-model parameters
% $U_0=2$, $\overline{U}=1$, $\zeta=0.8$, $D=0.5$, $\delta=0.1$,
% $\varepsilon_1=0.1$, $\varepsilon_2=0.2$, and $\sigma=0.5$, while $r$ is varied as the
% control parameter.

Let $(Q,U)$ denote the deterministic puff basic flow, governed by
Eq.~\eqref{eq:main_system} with the stochastic term omitted:
\begin{equation}
\begin{cases}
\dfrac{\partial Q}{\partial t}
+(U-\zeta)\dfrac{\partial Q}{\partial x}
=f(Q,U)+D_q\dfrac{\partial^2 Q}{\partial x^2},\\[0.8em]
\dfrac{\partial U}{\partial t}
+U\dfrac{\partial U}{\partial x}
=g(Q,U).
\end{cases}
\label{eq:puff_base}
\end{equation}
The basic flow $(Q,U)$ is a steadily propagating puff obtained from
Eq.~\eqref{eq:puff_base}.
We write the full fields as perturbations around this puff:
\begin{equation}
q_{\rm tot}=Q+q,\qquad
u_{\rm tot}=U+u .
\label{eq:perturbation_ansatz}
\end{equation}

Substituting Eq.~\eqref{eq:perturbation_ansatz} into Eq.~\eqref{eq:main_system} and subtracting Eq.~\eqref{eq:puff_base} gives the perturbation equations, which we write in residual form as
\begin{equation}
\left\{
\begin{aligned}
\mathcal{R}_q(q,u)
\equiv &\
\dfrac{\partial q}{\partial t}
+(U-\zeta)\dfrac{\partial q}{\partial x}
+u\dfrac{\partial Q}{\partial x}
+u\dfrac{\partial q}{\partial x}
-F(q,u)
-D_q\dfrac{\partial^2 q}{\partial x^2}
-\sigma (Q+q)\eta
=0,
\\[0.8em]
\mathcal{R}_u(q,u)
\equiv &\
\dfrac{\partial u}{\partial t}
+U\dfrac{\partial u}{\partial x}
+u\dfrac{\partial U}{\partial x}
+u\dfrac{\partial u}{\partial x}
-G(q,u)
=0,
\end{aligned}
\right.
\label{eq:residual_operators_barkley}
\end{equation}
where
\begin{equation}
F(q,u)=f(Q+q,U+u)-f(Q,U),\qquad
G(q,u)=g(Q+q,U+u)-g(Q,U).
\label{eq:FG_definition}
\end{equation}

% Throughout the optimization, the noise $\eta$ is treated as a fixed realization: the same realization that forces the direct
% equations~\eqref{eq:residual_operators_barkley} is stored along the
% direct trajectory and reused in the backward integration of the adjoint
% system derived below.

Equations~\eqref{eq:residual_operators_barkley}--\eqref{eq:FG_definition} constitute the full nonlinear evolution equations for the perturbation $(q,u)$ about the puff, and serve as the PDE constraint in the nonmodal optimization that follows.

\subsection{Masked nonlinear nonmodal analysis}

\subsubsection{The optimization procedure}
Nonlinear nonmodal analysis seeks the finite-amplitude initial perturbation of prescribed energy $E_0$ that maximizes a chosen objective over a finite time horizon $T$, thereby formulating transition as a nonlinear PDE-constrained optimization problem \cite{kerswell2014optimization,kerswell2018nonlinear}. In the present \emph{masked} formulation, the initial perturbation is additionally confined to a prescribed subregion of the domain, so that the receptivity of distinct parts of the puff can be probed separately.
% For $\sigma=0$, the formulation reduces to the standard deterministic analysis.

We take the disturbance kinetic energy of the total flow measured relative to
the laminar state $(q_{\rm tot},u_{\rm tot})=(0,U_0)$ at time $t=T$ as objective:
\begin{equation}
E_{k,T} =
\int
\frac{1}{2}
\left[
\left(q(x,T)+Q(x,T)\right)^2
+
\left(u(x,T)+U(x,T)-U_0\right)^2
\right]
\,\mathrm{d}x .
\label{eq:objective_functional_barkley}
\end{equation}
Maximizing $E_{k,T}$ thus favours initial perturbations that drive the flow
towards a sustained turbulent state, including two individual puffs evolving from single puff, rather than those producing only
transient amplification. The initial perturbation energy is prescribed through the masked constraint
\begin{equation}
E_0 =
\int
\frac{1}{2}
M(x)
\left[
q(x,0)^2+u(x,0)^2
\right]
\,\mathrm{d}x ,
\label{eq:initial_energy_constraint_barkley}
\end{equation}
where $E_0$ is a prescribed constant and $M(x)\ge 0$ is the mask that defines the target subregion. The initial perturbation is restricted to the support of the mask, i.e.\ $q(x,0)=u(x,0)=0$ wherever $M(x)=0$, and this restriction is maintained throughout the optimization, so that the constraint~\eqref{eq:initial_energy_constraint_barkley} controls the full initial perturbation energy.

Augmenting $E_{k,T}$ with the dynamic constraints~\eqref{eq:residual_operators_barkley} through the adjoint variables $p(x,t)$ and $v(x,t)$, and with the constraint~\eqref{eq:initial_energy_constraint_barkley} through the Lagrange multiplier $\lambda$, yields the Lagrangian
\begin{equation}
\begin{aligned}
\mathcal{L}
=&\
E_{k,T}
+
\lambda
\left[
\int
\frac{1}{2}
M(x)
\left(
q(x,0)^2+u(x,0)^2
\right)
\,\mathrm{d}x
-
E_0
\right]
\\
&+
\int_0^T
\int
p(x,t)\,
\mathcal{R}_q(q,u)
\,\mathrm{d}x\,\mathrm{d}t
\\
&+
\int_0^T
\int
v(x,t)\,
\mathcal{R}_u(q,u)
\,\mathrm{d}x\,\mathrm{d}t .
\end{aligned}
\label{eq:lagrangian_barkley}
\end{equation}
Let
\begin{equation}
F_q=\frac{\delta F}{\delta q},\quad
F_u=\frac{\delta F}{\delta u},\quad
G_q=\frac{\delta G}{\delta q},\quad
G_u=\frac{\delta G}{\delta u}.
\end{equation}
Taking variations of the Lagrangian
$\mathcal{L}$ with respect to $q$ and $u$ and integrating by parts in
space and time, where the spatial boundary terms vanish by periodicity,
yields the adjoint system:
\begin{equation}
\left\{
\begin{aligned}
&\frac{\partial p}{\partial t}
+(U+u-\zeta)\frac{\partial p}{\partial x}
+D_q\frac{\partial^2 p}{\partial x^2}
+\left(
\frac{\partial U}{\partial x}
+\frac{\partial u}{\partial x}
+F_q+\sigma\eta
\right)p
+G_q v
=0,
\\[0.8em]
&\frac{\partial v}{\partial t}
+(U+u)\frac{\partial v}{\partial x}
+G_u v
+\left(
F_u-\frac{\partial Q}{\partial x}
-\frac{\partial q}{\partial x}
\right)p
=0 .
\end{aligned}
\right.
\label{eq:adjoint_system}
\end{equation}

The terminal conditions are obtained from variations of $\mathcal{L}$
with respect to $q(x,T)$ and $u(x,T)$:
\begin{equation}
p(x,T)=-\left[q(x,T)+Q(x,T)\right],
\qquad
v(x,T)=-\left[u(x,T)+U(x,T)-U_0\right].
\label{eq:terminal_conditions}
\end{equation}

Variations of $\mathcal{L}$ with respect to $q(x,0)$ and $u(x,0)$ yield
the gradient
\begin{equation}
\frac{\delta\mathcal{L}}{\delta q(x,0)}=\lambda M(x)\,q(x,0)-p(x,0),
\qquad
\frac{\delta\mathcal{L}}{\delta u(x,0)}=\lambda M(x)\,u(x,0)-v(x,0),
\label{eq:initial_gradient}
\end{equation}
which provides the ascent direction for updating the initial perturbation under the masked energy constraint.
% At an optimum this gradient vanishes,
% giving the optimality conditions
% \begin{equation}
% p(x,0)=\lambda M(x)\,q(x,0),
% \qquad
% v(x,0)=\lambda M(x)\,u(x,0).
% \label{eq:optimality_conditions}
% \end{equation}

At the $i$-th ($i=0,1,2,...$) iteration, denote
\[
q_0^i=q^i(x,0),\quad u_0^i=u^i(x,0),\quad
p_0^i=p^i(x,0),\quad v_0^i=v^i(x,0).
\]
Then one ascent step is written as
\begin{equation}
\left\{
\begin{aligned}
q_0^{i+1}
&=
q_0^i
+\varepsilon
\left(
\lambda^i M q_0^i-p_0^i
\right),
\\[0.4em]
u_0^{i+1}
&=
u_0^i
+\varepsilon
\left(
\lambda^i M u_0^i-v_0^i
\right).
\end{aligned}
\right.
\label{eq:initial_gradient_update}
\end{equation}
Here $\varepsilon$ is the ascent step length. The multiplier $\lambda^i$
is determined by imposing the masked energy constraint on the updated
initial perturbation:
\begin{equation}
E_0
=
\int
\frac{1}{2}M
\left[
\left(q_0^{i+1}\right)^2
+
\left(u_0^{i+1}\right)^2
\right]
\,\mathrm{d}x .
\label{eq:lambda_energy_constraint}
\end{equation}
Substituting \eqref{eq:initial_gradient_update} into
\eqref{eq:lambda_energy_constraint} yields a quadratic equation for
$\lambda^i$,
\begin{equation}
A(\lambda^i)^2+B\lambda^i+C=0,
\label{eq:lambda_quadratic}
\end{equation}
where
\begin{equation}
\begin{aligned}
A
&=
\frac{\varepsilon^2}{2}
\int
M^3
\left[
(q_0^i)^2+(u_0^i)^2
\right]
\,\mathrm{d}x,
\\
B
&=
\varepsilon
\int
M^2
\left[
q_0^i(q_0^i-\varepsilon p_0^i)
+
u_0^i(u_0^i-\varepsilon v_0^i)
\right]
\,\mathrm{d}x,
\\
C
&=
\frac{1}{2}
\int
M
\left[
(q_0^i-\varepsilon p_0^i)^2
+
(u_0^i-\varepsilon v_0^i)^2
\right]
\,\mathrm{d}x
-E_0 .
\end{aligned}
\label{eq:lambda_ABC}
\end{equation}
Thus,
\begin{equation}
\lambda^i
=
\frac{-B\pm\sqrt{B^2-4AC}}{2A}.
\label{eq:lambda_solution}
\end{equation}
Both roots satisfy the energy
constraint~\eqref{eq:lambda_energy_constraint} by construction. We choose
between them from the small-step limit. Suppose the current iterate
already satisfies the constraint. The coefficients in~\eqref{eq:lambda_ABC}
then scale as $A \sim O(\varepsilon^2)$, $B \sim O(\varepsilon)$ and
$C \sim O(\varepsilon)$. Here $A>0$ because the mask $M$ is non-negative, and
$B>0$ for small $\varepsilon$. The discriminant $B^2-4AC$ is then
positive, because $B^2 \sim O(\varepsilon^2)$ dominates $4AC \sim O(\varepsilon^3)$,
so both roots are real. They split into a regular branch and a singular
branch. The regular branch
$\lambda^i\to-C/B$ has the finite limit
\[
\lambda^i\to
\frac{\int M\bigl(q_0^i p_0^i+u_0^i v_0^i\bigr)\,\mathrm{d}x}
{\int M^2\bigl[(q_0^i)^2+(u_0^i)^2\bigr]\,\mathrm{d}x},
\qquad\varepsilon\to0 .
\]
The singular branch $\lambda^i\sim-B/A$ diverges,
\[
\lambda^i\sim
-\frac{2}{\varepsilon}\,
\frac{\int M^2\bigl[(q_0^i)^2+(u_0^i)^2\bigr]\,\mathrm{d}x}
{\int M^3\bigl[(q_0^i)^2+(u_0^i)^2\bigr]\,\mathrm{d}x}
\to-\infty .
\]
On the singular branch $\lambda^i \sim O(\varepsilon^{-1})$, so the update
changes the initial condition by $O(1)$ and does not vanish as
$\varepsilon\to0$. This is not a small step, so we keep the regular
branch. For small
$\varepsilon$ it is the larger of the two roots, that is the $+$ sign
in~\eqref{eq:lambda_solution}. The update~\eqref{eq:initial_gradient_update}
is then small, $q_0^{i+1}-q_0^i \sim O(\varepsilon)$ and
$u_0^{i+1}-u_0^i \sim O(\varepsilon)$, and reduces to the identity as
$\varepsilon\to0$.

\begin{figure}
    \centering
    \includegraphics[width=0.8\linewidth]{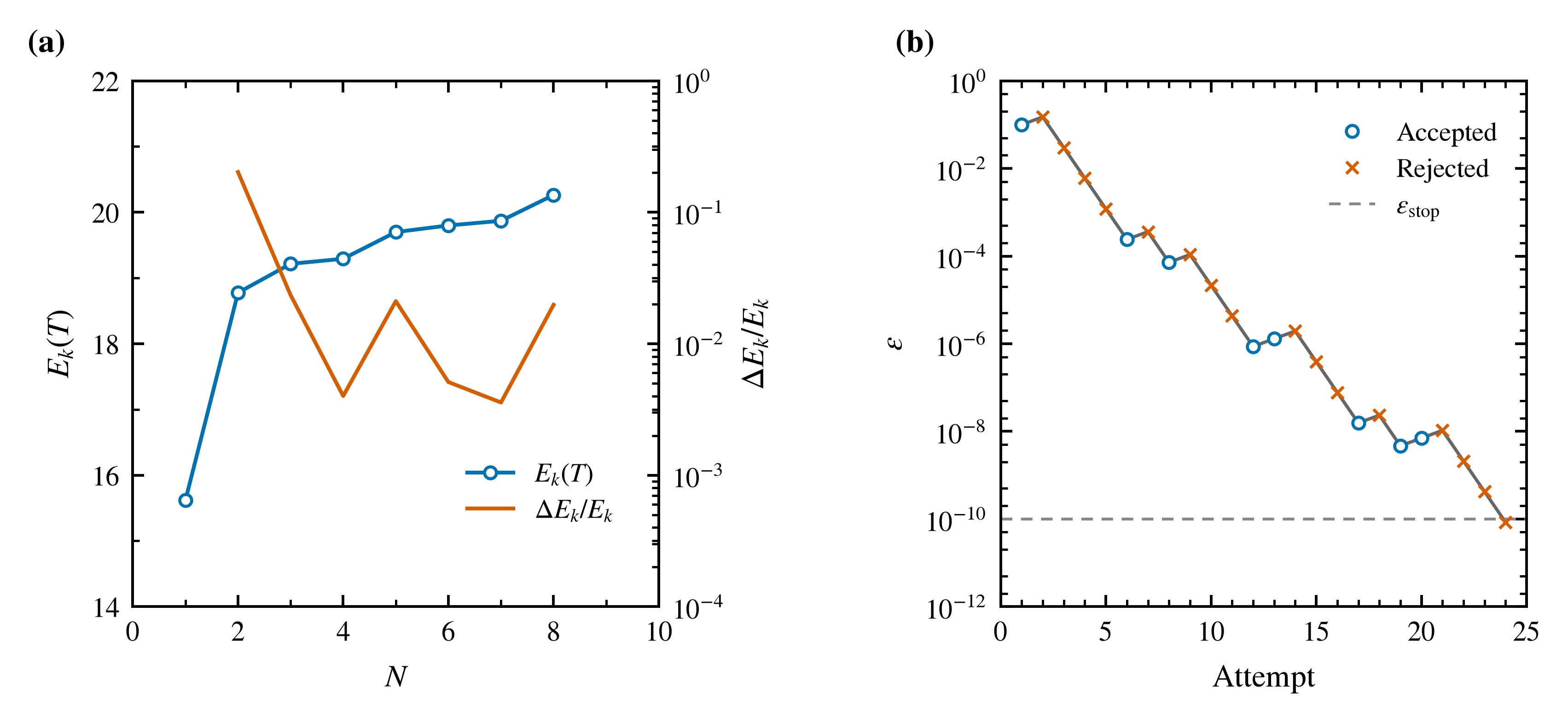}
    \caption{Key optimization metrics during NNO. (a) shows the terminal energy $E_k(T)$ (blue open circles) and its relative increase between successive accepted optimization steps (solid orange line) for the noiseless optimization. The horizontal axis $N$ represents the number of accepted optimization iteration steps. (b) shows the evolution of the optimization step-size parameter $\epsilon$ over all attempts. Blue open circles and orange crosses indicate accepted and rejected attempts, respectively. The dashed horizontal line marks the stopping threshold $\varepsilon_{\mathrm{stop}}=10^{-10}$ applied to rejected attempts.}
    \label{fig:NNOenergy}
\end{figure}

The nonlinear nonmodal optimization process is as follows.
The perturbation is initialized as random
noise supported on $M$ and normalized to the masked energy
$E_0$~\eqref{eq:initial_energy_constraint_barkley}.
Each iteration then
proceeds in four stages:
(i) the direct
system~\eqref{eq:residual_operators_barkley} is integrated forward over
$[0,T]$ at fixed noise;
(ii) the adjoint terminal state $(p,v)|_{t=T}$ is
set from~\eqref{eq:terminal_conditions}; (iii) the adjoint
system~\eqref{eq:adjoint_system} is integrated backward to $t=0$;
and (iv) the initial condition is updated
through~\eqref{eq:initial_gradient_update}, with $\lambda^i$ fixed
by~\eqref{eq:lambda_solution}.
The loop is iterated until $E_{k,T}$ converges to
within a prescribed tolerance, just as shown in Fig.~\ref{fig:NNOenergy}. Here when the relative error $\Delta E_k/E_k \leq 10^{-2}$ or or $\varepsilon \leq \varepsilon_{\mathrm{stop}}=10^{-10}$, the optimization results are checked by DNS until the flow field with two puffs could persist for a long time.
% \begin{figure}
%     \centering
%     \includegraphics[width=0.8\linewidth]{figures/NNOenergy.png}
%     \caption{(a) and (b) display the convergence curves for optimization with and without the noise term, respectively. Blue circles denote the disturbance kinetic energy of the flow field at target time $T$, while the solid orange line represents the relative error between two successive optimization steps.}
%     \label{fig:NNOenergy}
% \end{figure}

For the noisy optimization, we employ a warm-start strategy. Noise enhances trajectory variability and the sensitivity of the optimization to the initial guess, making direct optimization over long horizons more challenging\citep{Monokrousos2011,kerswell2018nonlinear}. We therefore increase the optimization horizon sequentially through $T=20,40,60,80,100$, using the optimized initial perturbation at each stage to initialize the next. The initial perturbation energy remains fixed, and each trajectory is integrated anew from $t=0$ with the same noise realization over overlapping intervals. The disturbance kinetic energy evolution of intermediate optimization steps with different target time $T$ is shown in Fig.~\ref{fig:NNO}(c) . The transferred field is thus the optimized initial perturbation, rather than the terminal state. This warm-start procedure facilitates the search for perturbations that induce splitting at longer horizons.

\subsubsection{Comparison of noisy and noiseless optimization}
\label{sec:noise_noiseless_comparison}

\begin{figure}[t]
    \centering
    \includegraphics[width=0.5\linewidth]{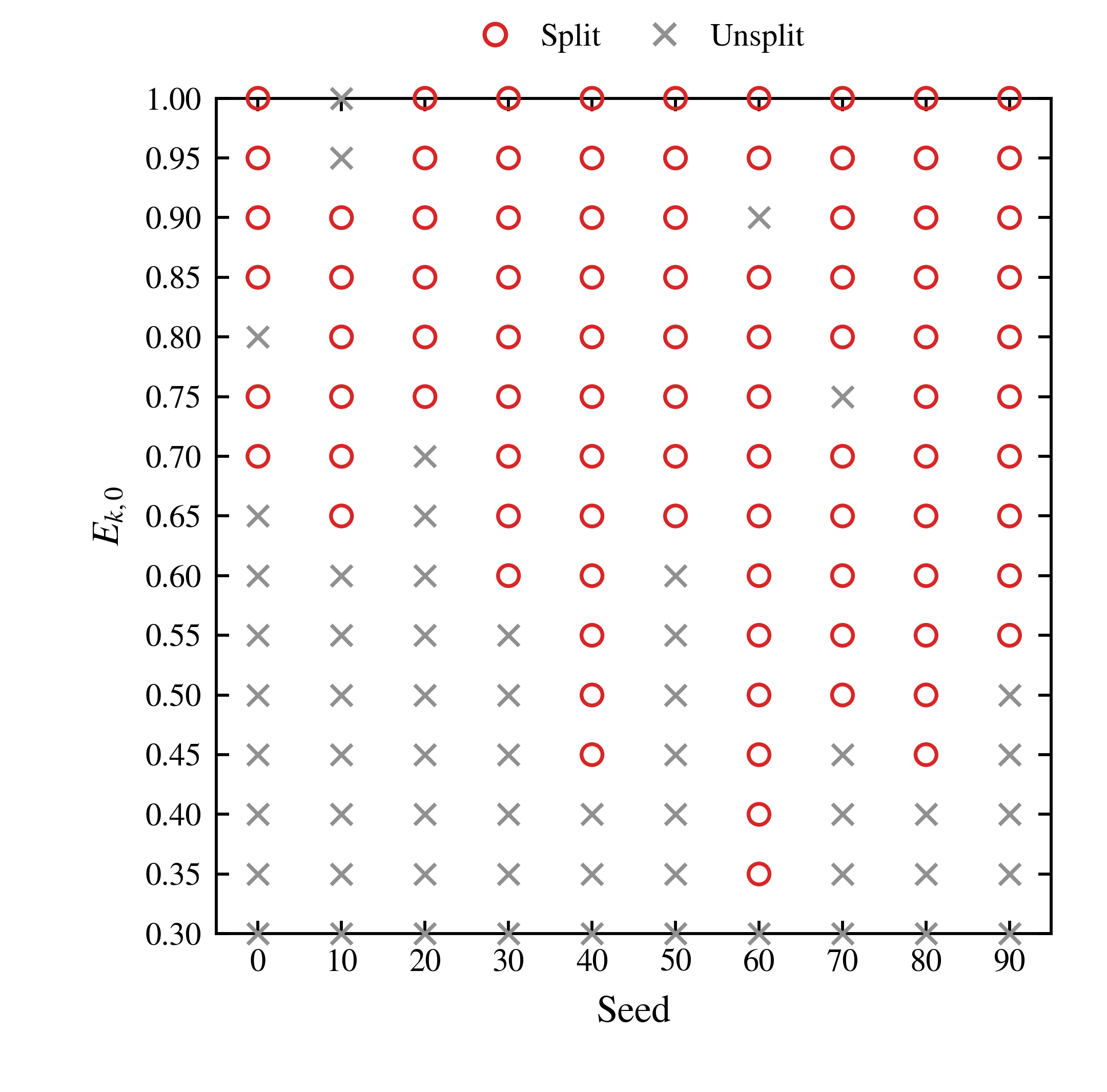}
    \caption{Splitting phase diagram for noisy optimization at $r=0.7$ and $\sigma=0.5$. The horizontal axis denotes the random seed identifying each noise realization, and the vertical axis denotes the initial perturbation energy $E_{k,0}$. The optimization horizon is increased sequentially through $T=20,40,60,80,100$. Red open circles and gray crosses indicate splitting and nonsplitting outcomes at $T=100$, respectively.}
    \label{fig:noise_seed_energy_phase}
\end{figure}

Both noisy and noiseless optimization maximize the terminal total-field energy at fixed initial perturbation energy to identify perturbations that induce puff splitting. The noisy formulation additionally includes stochastic forcing in the evolution equations. Each noise realization is held fixed throughout the optimization, while varying the random seed allows us to assess the dependence of the outcome on the forcing realization.

Figure~\ref{fig:noise_seed_energy_phase} shows that increasing the initial perturbation energy generally promotes splitting for a fixed noise realization. The lowest observed energy triggering splitting is approximately $E_{k,0}=0.35$, whereas a substantial fraction of the sampled noise realizations exhibit splitting only at energies around 0.5 and above. These results are qualitatively consistent and in quantitative agreement with the energy thresholds and phase diagram in Fig. 4(a). The small differences likely reflect the non-negligible noise and the staged warm-start optimization used here, which can identify splitting-inducing perturbations at somewhat lower initial energies.
These results demonstrate that both approaches can identify splitting-inducing perturbations, although noisy optimization does not yield splitting for every noise realization.

% These findings do not establish that the two approaches have identical splitting thresholds or search success rates. Moreover, noiseless optimization is performed directly at $T=100$, whereas noisy optimization uses horizon continuation. The observed differences therefore cannot be attributed solely to noise. A quantitative comparison requires consistent optimization protocols and splitting criteria, together with a systematic comparison of the corresponding phase diagrams.

\subsection{Numerical simulations}
\subsubsection{Simulations of stochastic differential equation}
% NH：以下注释内容来自于我此前文档的数值部分，师兄如有需要可以使用一部分
To solve the forward stochastic perturbation equations, we employ the Euler--Maruyama (EM) discretization. Since the stochastic perturbation equation is interpreted in the It\^o sense, its $q$ equation can be written as:
\begin{equation}
\mathrm{d}q =  N_q(q,u)  \mathrm{d}t + \sigma (Q+q) \mathrm{d}W.
\end{equation}
Here, $W$ denotes the stochastic driving process. Let $\Delta t$ and $\Delta x$ be the temporal and spatial step sizes. The discrete white noise increment $\Delta W_j^n$ at grid point $x_j$ and time $t_n$ is sampled as:
\begin{equation}
\Delta W_j^n = \sqrt{\Delta t} \chi_j^n, \quad \chi_j^n \sim \mathcal{N}(0,1),
\end{equation}
where $\chi_j^n$ are independent and identically distributed (i.i.d.) standard normal random variables.
% The division by $\sqrt{\Delta x}$ ensures that the spatial variance correctly approximates the delta-correlation of space-time white noise.

The update step for the perturbation variables $(q_j^n, u_j^n) \to (q_j^{n+1}, u_j^{n+1})$ is given by:
\begin{align}
q_j^{n+1} &= q_j^n +  N_q(q_j^n, u_j^n)  \Delta t + \sigma (Q_j^n+q_j^n) \Delta W_j^n, \\
u_j^{n+1} &= u_j^n + N_u(q_j^n, u_j^n) \Delta t,
\end{align}
where $N_q$ and $N_u$ represent the deterministic spatial operators (convection, diffusion, and reaction). For the adjoint backward integration, the corresponding discrete adjoint step is performed for $p$ using the same realization of $\Delta W_j^n$ stored during the forward pass.

All computations are performed in a one-dimensional periodic domain of length $L=100$ with $N_x=1000$. Spatial derivatives are evaluated using a Fourier pseudo-spectral method. The base-state traveling puff is pre-computed until convergence (residual $< 5\times 10^{-5}$). 

The optimization parameters are:
\begin{itemize}
    \item Model parameters: $r=0.7, \zeta=0.8, \bar{U}=1, U_0=2, D_q=0.5, \delta=0.1, \epsilon_1=0.1, \epsilon_2=0.2$.
    \item Optimization: $E_0=0.5, T=100$.
    \item Noise intensity: $\sigma = 0.5$. 
    % (prescribed per case).
    \item Time step: $\Delta t=0.01$. 
\end{itemize}

The initial mask $M(x)$ is defined based on the puff energy $\mathcal{E}_{\mathrm{puff}}(x) = \frac{1}{2}[Q(x,0)^2+(U(x,0)-U_0)^2]$:
\begin{equation}
M(x)=
\begin{cases}
1, &\{ x | \frac{\mathcal{E}_{\mathrm{puff}}(x)}{\mathop{max}\limits_{x} \mathcal{E}_{\mathrm{puff}}(x)}>0.01 \ \cap \ x>x_{\max} \},\\
0, & \text{otherwise},
\end{cases}
\end{equation}
where $x_{\max}$ is the peak position of $\mathcal{E}_{\mathrm{puff}}(x)$. This restricts perturbations to the downstream "active" region of the puff. The gradient ascent is performed with an adaptive step size $\varepsilon$, increasing by $1.5$ upon a successful cost increase and decreasing by $0.2$ upon failure.

\subsubsection{Direct numerical simulations of Navier-Stokes equations}\label{app:DNS}

Direct Numerical Simulations (DNS) of the incompressible pipe flow governed by the Navier-Stokes equations were performed using OpenPipeflow \cite{willis2017}. A constant flow rate, corresponding to a fixed bulk velocity $U$, was imposed to maintain a temporally fixed Reynolds number $Re=2230$. Periodic boundary conditions were applied in the streamwise direction in a domain of length $L=32\pi D\approx100D$, while no-slip boundary conditions were imposed at the pipe wall. The velocity and pressure fields were spatially discretized in cylindrical coordinates $(x,\rho,\theta)$ using a hybrid spectral finite-difference method, with $\pm768$ axial Fourier modes, $52$ radial points located at the roots of a Chebyshev polynomial, and $\pm32$ azimuthal Fourier modes. The resolution used in the OpenPipeflow code here is slightly higher than that used and validated in previous studies \cite{WillisKerswell2007, avila2011, Svirsky2026, NemotoAlexakis2021} at Reynolds numbers where puff splitting is observed. Time integration was carried out using a second-order predictor-corrector scheme with a time step of $0.0025D/U$. The initial conditions were randomly selected velocity snapshots at $Re=2040$ from a previous study \cite{guan2025}, containing quasi-equilibrium puff states. Profiles of $u_\mathrm{DNS}$ and $q_\mathrm{DNS}$ were saved in a time interval of $0.05 D/U_b$. 

% The survivor function $S(t)=1-\operatorname{CDF}(t)$, where $\operatorname{CDF}(t)$ is the Cumulative Distribution Function, is shown in Fig.~\ref{fig:time}. The lifetimes are fitted by an exponential distribution, whose slope on the semi-log axis represents the mean value. The mean lifetime at $Re=2230$ is compared with existing experimental data and fitted scaling \cite{avila2011} for validation, as shown in Fig.~\ref{fig:time}. 

% \begin{figure}[h]
%     \centering
%     \includegraphics[width=0.4\linewidth]{figures/time.jpg}
%     \caption{Survivor function of puffs at $Re=2230$. Results of individual lifetimes are shown by circular symbols. The DNS results are compared with existing experimental data and fitted scaling of mean lifetimes \cite{avila2011}.}
%     \label{fig:time}
% \end{figure}

\subsection{Profile matching and ensemble averaging}
\label{sec:dns_registration}

We compare DNS profiles with the noiseless NNO evolution at $r=0.7$ and $T=100$. Superscripts $\mathrm{DNS}$ and $\mathrm{NNO}$ distinguish the two data sets. Spatial registration separates the internal shape of a localized structure from its position and overall width. Following the center and length definition of the localized wave packet in previous works \citep{ZhangTao2022Pattern}, we also define the puff center $x_c$ and length $l_p$ by
\begin{equation}
\begin{aligned}
x_c&=\frac{\int xw\,\mathrm{d}x}{\int w\,\mathrm{d}x},\\
l_p^2&=\frac{12\int(x-x_c)^2w\,\mathrm{d}x}{\int w\,\mathrm{d}x}.
\end{aligned}
\label{eq:dns-moments-en}
\end{equation}
The weight is $w=(u_{\rm tot}^{\mathrm{DNS}}-2)^2$ for DNS and $w=(u_{\rm tot}^{\mathrm{NNO}}-U_0)^2$ for NNO. This nonnegative quadratic measure emphasizes the departure from laminar flow. Normalized moments are independent of its overall amplitude.

% All spatial operations use the periodic domain $[-50,50)$. The maximum of $w$ is first shifted to the domain center, placing the localized structure away from the periodic cut. Identical Fourier translations are then applied to both total fields until $|x_c|<10^{-7}$. DNS profiles are periodically interpolated onto the 512-point NNO grid, recentered, and assigned their lengths. For each pair $(t_{\mathrm{DNS}},t_{\mathrm{NNO}})$, NNO profiles are sampled at $x\ell_{\mathrm{NNO}}/\ell_{\mathrm{DNS}}$ to match the instantaneous DNS width. Linear interpolation is used within the sampled interval, with laminar continuation $(q_{\rm tot},u_{\rm tot})=(0,U_0)$ outside it. This continuation preserves a zero-disturbance far field under the assumption of localized tails. The scaled NNO fields are then jointly recentered using the same weight.
% The aligned DNS and NNO profiles are then jointly mapped back onto the length scale of the corresponding DNS profile through an inverse rescaling. This ensures that both profiles are represented on the same DNS-based spatial scale, allowing the subsequent ensemble average to retain a clear physical interpretation as a length-conditioned ensemble.
All spatial operations use the periodic domain $[-50,50)$. The maximum of $w$ is first shifted to the domain center, followed by identical Fourier translations of both total fields until $|x_c|<10^{-7}$. DNS profiles are periodically interpolated onto the 512-point NNO grid, recentered, and their lengths evaluated. For calculating $I$, NNO profiles are sampled at $xl_{p,\mathrm{NNO}}/l_{p,\mathrm{DNS}}$ using linear interpolation with laminar continuation $(q_{\rm tot},u_{\rm tot})=(0,U_0)$, then recentered. This length matching is used only for similarity-based time selection.

In the overlap below, total-field symbols denote these registered profiles. With $\langle f,g\rangle=\int fg\,\mathrm{d}x$ and $\|f\|^2=\langle f,f\rangle$, define
\begin{equation}
\begin{aligned}
I(t_{\mathrm{DNS}};t_{\mathrm{NNO}})
={}&\frac12
\frac{\langle q_{\rm tot}^{\mathrm{DNS}},q_{\rm tot}^{\mathrm{NNO}}\rangle}
{\|q_{\rm tot}^{\mathrm{DNS}}\|\,\|q_{\rm tot}^{\mathrm{NNO}}\|}\\
&+\frac12
\frac{\langle u_{\rm tot}^{\mathrm{DNS}}-2,u_{\rm tot}^{\mathrm{NNO}}-U_0\rangle}
{\|u_{\rm tot}^{\mathrm{DNS}}-2\|\,\|u_{\rm tot}^{\mathrm{NNO}}-U_0\|}.
\end{aligned}
\label{eq:dns-overlap-en}
\end{equation}
Moments and overlaps use trapezoidal quadrature on the common sampled interval. Subtracting the laminar velocity isolates the localized departure, while separate normalization gives both component shapes equal weight. For nonzero norms, $I=1$ corresponds to positive proportionality of each registered component pair.

Matching proceeds at $t_{\mathrm{NNO}}=0,10,\ldots,100$. The first search covers saved DNS snapshots in last 400 $t_\mathrm{DNS}$.
Each subsequent search is restricted to times later than the preceding match. Within each remaining window, the empirical 99th percentile of $I$ is recomputed by linear interpolation. The selected time is the last local maximum in the last contiguous interval above or equal to this threshold. A peak is at least as high as both neighbors in the full sequence and strictly higher than one. The interval maximum provides the fallback, with ties resolved by the earliest sample. This rule selects late, high-overlap events while preserving temporal order and allowing different evolution rates. The selected times for the DNS trajectories are listed in Table~\ref{tab}.

The ensembles are formed from the original DNS fields at the selected times. At each reference time, the ensemble average is computed from these aligned DNS fields using normalized peak-similarity weights $\omega_i=I_i/\sum_j I_j$.

\begin{table*}[t]
\caption{Selected DNS times $t_\mathrm{DNS}$ corresponding to the NNO reference states at $t_\mathrm{NNO}=0,10,...,100$ for 11 DNS cases. Matching times are identified from the similarity measure $I$ using a sequential peak-selection procedure based on the 99th percentile, with each selected time strictly later than the preceding one. Rows denote DNS cases, and columns denote NNO reference times. Cases 8--11 are used to compute the ensemble averages shown in Fig.~3(b)--(e) of the main text.}
\centering
\small
\setlength{\tabcolsep}{3pt}
\renewcommand{\arraystretch}{1.15}
\begin{tabular}{l*{11}{r}}
\hline\hline
 & \multicolumn{11}{c}{$t_{\mathrm{NNO}}$} \\
\cline{2-12}
% s12,s16,s17,s18,s19,s24,s25,xu4,xu5,xu6,xu7
DNS & 0 & 10 & 20 & 30 & 40 & 50 & 60 & 70 & 80 & 90 & 100 \\
\hline
case1 & 2551.80 & 2657.60 & 2658.50 & 2662.25 & 2662.95 & 2666.20 & 2713.80 & 2714.30 & 2720.90 & 2721.60 & 2727.30 \\
case2 & 2964.80 & 3097.00 & 3103.20 & 3103.60 & 3107.60 & 3108.40 & 3109.85 & 3113.60 & 3136.95 & 3139.20 & 3139.25 \\
case3 & 2931.85 & 2987.60 & 3041.85 & 3042.70 & 3045.60 & 3047.70 & 3048.85 & 3089.45 & 3104.35 & 3122.45 & 3122.50 \\
case4 & 4729.75 & 4894.60 & 4903.20 & 4948.85 & 4950.65 & 4956.55 & 4959.05 & 4960.85 & 4974.20 & 5009.90 & 5011.05 \\
case5 & 2252.25 & 2339.00 & 2340.55 & 2342.00 & 2343.45 & 2400.75 & 2400.80 & 2400.85 & 2407.55 & 2407.75 & 2407.90 \\
case6 & 5845.35 & 5938.30 & 5967.85 & 5969.20 & 5972.00 & 5973.65 & 6051.65 & 6052.35 & 6073.65 & 6073.80 & 6073.85 \\
case7 & 1376.40 & 1489.90 & 1493.40 & 1507.70 & 1528.00 & 1530.75 & 1536.00 & 1542.20 & 1577.45 & 1633.75 & 1760.20 
\\
case8 & 2564.00 & 2660.85 & 2709.65 & 2737.75 & 2748.95 & 2783.95 & 2786.85 & 2797.50 & 2817.30 & 2820.50 & 2874.55 \\
case9 & 716.40 & 847.15 & 865.90 & 905.65 & 910.05 & 914.35 & 915.50 & 1092.15 & 1092.55 & 1092.60 & 1092.65 
\\
case10 & 7283.20 & 7373.45 & 7374.55 & 7415.65 & 7419.15 & 7420.25 & 7420.60 & 7420.65 & 7430.70 & 7435.35 & 7439.45 
\\
case11 & 1590.80 & 1663.90 & 1675.30 & 1681.90 & 1689.40 & 1698.20 & 1700.65 & 1707.70 & 1842.90 & 1906.20 & 1906.25 \\
\hline\hline
\end{tabular}
\label{tab}
\end{table*}

% For display, the matched DNS profiles and the corresponding NNO profile are mapped to the mean matched DNS length. Each profile is sampled at $x$ multiplied by its own length divided by this mean, continued by its respective laminar values, and recentered. The two DNS components are then multiplied separately by the ratio of the NNO peak to their own peak. Peaks are evaluated over the full grid, with the centered NNO before display scaling as reference. The velocity gain acts on the total $u_{\rm tot}$, including its background. This amplitude matching follows time selection, and the ensemble is the pointwise arithmetic mean of the resulting profiles. Registration before averaging retains the common morphology by removing displacement and width variability. The resulting mean describes the amplitude-normalized shape conditioned on the matching rule.

\end{document}